\documentclass[twoside]{LCWS11}
\usepackage[latin1]{inputenc}
\usepackage[dvips]{graphicx,epsfig,color}
\usepackage{subfigure}
\usepackage{wrapfig,rotating}
\usepackage{amssymb,amsmath,array}
\usepackage{cite}

\begin{document}
\title{Recent Advances of the Engineering Prototype of the CALICE
  Analog Hadron Calorimeter} 
\author{Oskar Hartbrich$^1$ and Mark Terwort$^2$ on behalf of the CALICE collaboration
\vspace{.3cm}\\
1- Bergische Universit\"at Wuppertal \\
Gau{\ss}stra{\ss}e 20, 42119 Wuppertal - Germany
\vspace{.1cm}\\
2- DESY \\
Notkestra{\ss}e 85, 22607 Hamburg - Germany\\
}

\maketitle

\begin{abstract}

  The CALICE collaboration is developing an engineering prototype of
  an analog hadron calorimeter for a future linear collider detector.
  The prototype has to prove the feasibility of building a realistic
  detector with fully integrated front-end electronics. The
  performance goals are driven by the requirement of high jet energy
  resolution and the measurement of the details of the shower
  development. The signals are sampled by small scintillating plastic
  tiles that are read out by silicon photomultipliers. The ASICs are
  integrated into the calorimeter layers and are optimized for minimal
  power consumption. For the photodetector calibration an LED system
  is integrated into each of the detector channels. In this report the
  status and performance of the realized module are presented. In
  particular, results from timing measurements are discussed, as well
  as tests of the calibration system. The new module has also been
  used in the DESY test beam environment and first results from the
  electron beam tests are reported.

\end{abstract}

\section{Introduction} \label{}

\begin{wrapfigure}{r}{0.4\columnwidth}
\centerline{\includegraphics[width=0.35\columnwidth]{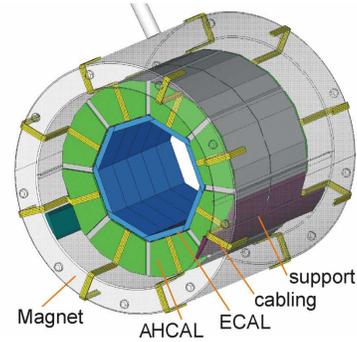}}
\caption{Layout of the barrel calorimeter system of a LC detector. The
  AHCAL is shown in green and the ECAL in blue, while the structure is
  surrounded by the magnet.}
\label{fig:barrel}
\end{wrapfigure}

Within the CALICE collaboration~\cite{CALICE} new technologies for
calorimeters for a future linear collider (LC) experiment are
developed and tested. Figure~\ref{fig:barrel} shows a possible design
of the (barrel) calorimeters for a LC detector. The sandwich analog
hadron calorimeter (AHCAL) with 48 layers is a cylindrical structure
with an inner and outer radius of 2.0\,m and 3.1\,m, respectively.
Inside the AHCAL the electromagnetic calorimeter (ECAL) will be
placed, while it is surrounded by the magnet. A major aspect for the
design is the improvement of the jet energy resolution compared to
previous experiments. This can be achieved by measuring the details of
the spatial shower development for a good shower separation and
combining these information with measurements from the tracking
detectors. This approach is known as {\it particle flow} and has been
validated~\cite{PF} with the physics prototype~\cite{PPT} of the
CALICE AHCAL. A very high segmentation of the calorimeters in all
dimensions is mandatory for a good performance of particle flow
algorithms.

A new engineering prototype~\cite{EPT} is currently being developed to
demonstrate that a scalable device can be built that meets the
requirements of an LC experiment. Key requirements for the front-end
electronics are very low power consumption and full integration into
the active calorimeter layers. The prototype is based on scintillating
tiles that are read out by silicon photomultipliers (SiPMs). First
subunits (HCAL base unit, HBU) with 144 detector channels of size
$36\times 36$\,cm$^2$ have been designed and extensively tested in the
laboratory as well as in the DESY test beam facility.

In Sec.~\ref{sec:status} the concept and current status of the
prototype are presented, while results from recent tests are discussed
in Sec.~\ref{sec:results}.

\section{Design and status of the engineering prototype} \label{sec:status}

\begin{wrapfigure}{r}{0.4\columnwidth}
\centerline{\includegraphics[width=0.3\columnwidth]{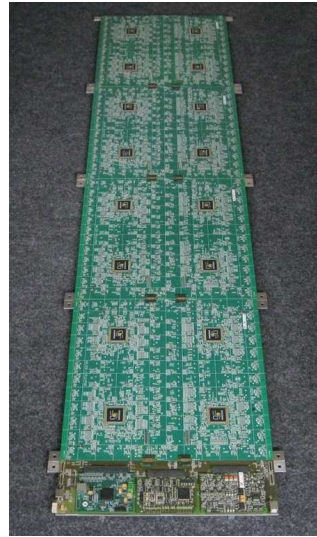}}
\caption{Photo of four assembled HBUs. The detector interface modules
  are also shown.}
\label{fig:slab}
\end{wrapfigure}

Figure~\ref{fig:barrel} illustrates that the AHCAL is divided into two
sections along the beam direction and in 16 sectors in
$\phi$-direction. Each sector consists of 48 layers with a total
thickness of 110\,cm and a length of 220\,cm. A single layer consists
of 16\,mm thick stainless steel (or 10\,mm thick tungsten) absorber
plates and an active layer part that is subdivided into several HBUs.
Each layer has about 2500 channels, which adds up to about 4 million
channels for the whole barrel AHCAL. All electronics connections and
interface modules are placed at the two end-faces of the barrel, which
are easily accessible for maintenance and service lines.

\begin{figure}
\centering
\subfigure[] {\includegraphics[width=2.5in]{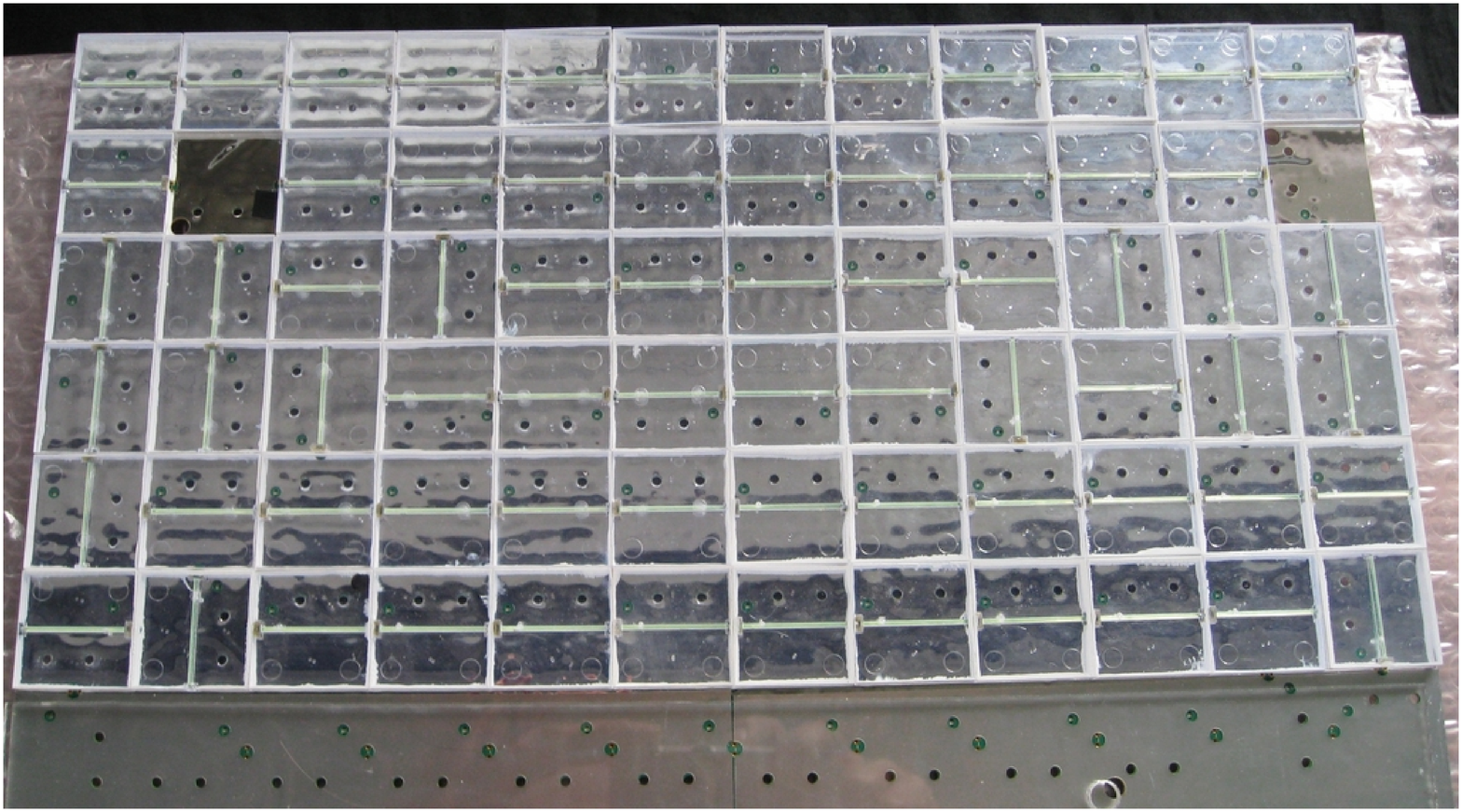}\label{fig:tiles}}
\hspace{1.5cm}
\subfigure[] {\includegraphics[width=1.5in]{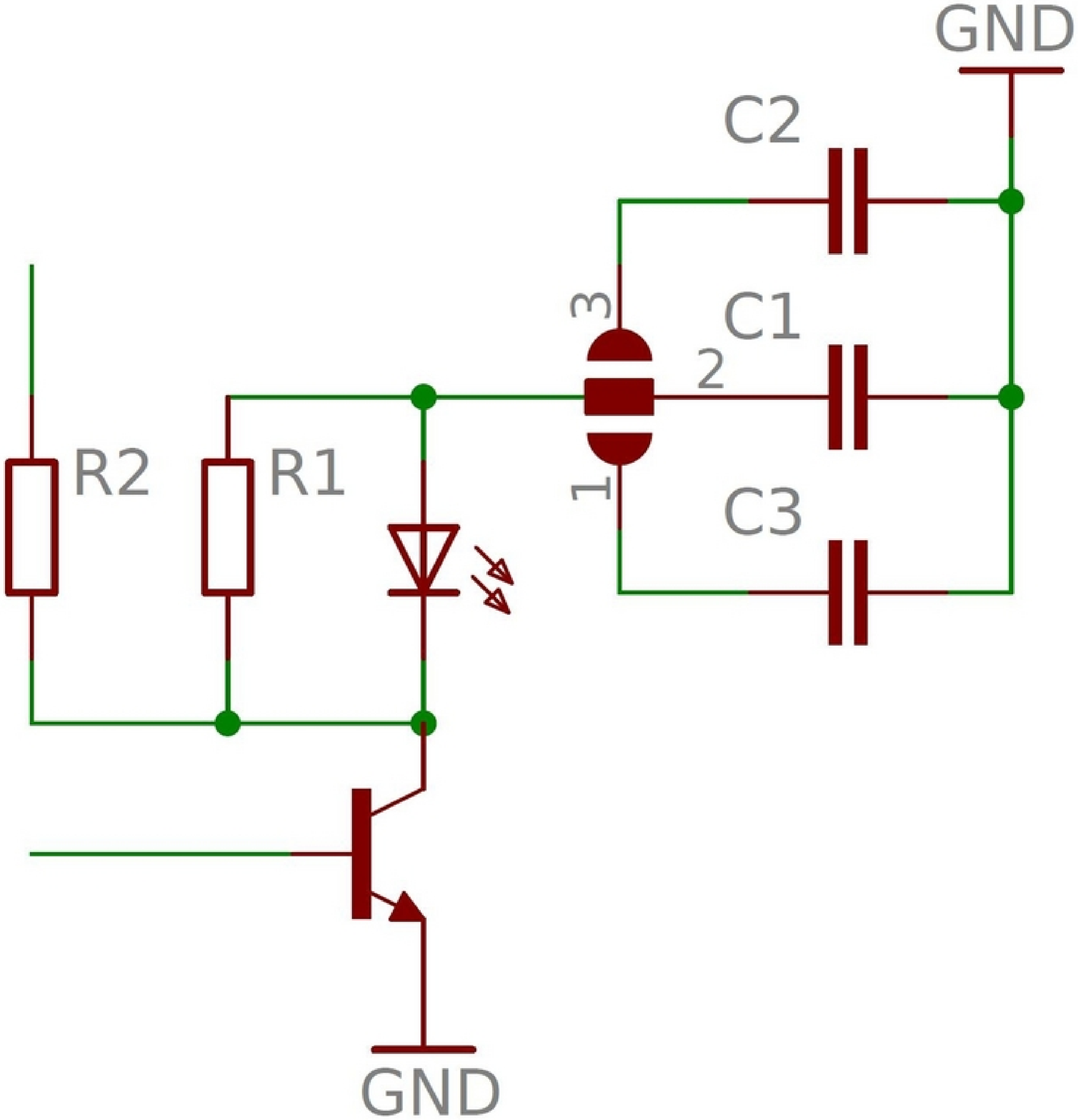}\label{fig:LED}}
\caption{(a) Photo of 70 tiles that are assembled below an HBU. The
  orientation of the wavelength shifting fibers depends on the
  position of the holes for the pins. This is determined by the
  details of the PCB design. (b) Diagram of the LED driver circuit.}
\end{figure}

The AHCAL layers are subdivided into three parallel slabs, which
consist of six HBUs that are interconnected via ultra-thin flex leads.
Figure~\ref{fig:slab} shows the current setup of four assembled HBUs.
Each HBU features 144 detector channels of $3\times 3$\,cm$^2$ size.
The energy deposited in the calorimeter is sampled by 3\,mm thick
scintillating plastic tiles and the scintillation light is guided with
an integrated wavelength shifting fiber to a SiPM with a size of
1.27\,mm$^2$. The SiPMs comprise 796 pixels operated in Geiger mode
with a gain of $\sim 0.5\cdot 10^6$ - $2.0\cdot 10^6$. The tiles are
connected below the PCB with a nominal distance of 100\,$\mu$m by two
alignment pins that are plugged into holes in the PCB. A photo of 70
assembled tiles on the backside of an HBU is shown in
Fig.~\ref{fig:tiles}.

The 36-channel SPIROC2b ASICs~\cite{ASICs}, that read out the analog
signals from the SiPMs, are mounted on the top side of the PCB and are
lowered into cutouts by $\sim$500\,$\mu$m to reduce the height of the
active layers. They are equipped with 5\,V DACs for channel-wise bias
voltage adjustment. Two gain modes are provided, where the high gain
mode is primarily foreseen for taking calibration data and the low
gain mode measures signals with higher amplitudes up to SiPM
saturation. To avoid the need for an active cooling system inside the
calorimeter layers for the final LC operation, the power consumption
has to be limited to 25\,$\mu$W (40\,$\mu$W) per channel (including
SiPMs). This is only possible when parts of the ASICs are switched off
when they are not needed. This is done according to the bunch train
structure of the LC ({\it power pulsing})~\cite{Peter}.  The
on-detector zero suppression with an adjustable threshold is
integrated into the ASICs as well as the digitization step with a
12-bit ADC for charge and a 12-bit TDC for time measurements. The TDC
comprises two ramps with variable lengths between 200\,ns and
5\,$\mu$s, depending on the operation mode (LC or test beam). The
working principle as well as the performance of the TDC in LC and test
beam mode are discussed in Sec.~\ref{sec:TDC}.

Since the response of SiPMs strongly depends on the temperature
($\sim$-1.7\%/K) and the applied bias voltage ($\sim$2.5\%/100\,mV), a
calibration system is needed in order to correct for these effects.
Furthermore, SiPMs saturate at high light intensities due to the
limited number of pixels, which also has to be measured. Therefore,
each channel contains a circuit for pulsing an integrated UV LED,
where the light amplitude can be controlled by an external voltage. A
fast trigger with a pulse width of 40\,ns and an analog bias voltage
of 0\,V - 10\,V is provided by the corresponding detector interface
module (middle interface board in Fig.~\ref{fig:slab}). For the gain
calibration low light intensities are used to extract the gain from
the distances of individual peaks in a single-pixel-spectrum, while
for higher light intensities (corresponding to $\sim$100
minimum-ionizing particles) the SiPMs saturate. Figure~\ref{fig:LED}
shows a circuit diagram of the LED driver. The middle bias capacitor
C$_1=150$\,pF is charged and upon an LED trigger signal, that opens
the transistor, C$_1$ is discharged by a current flowing through the
LED. The resistor R$_1=50\,\Omega$ guarantees a fast fall time of the
optical LED pulse. Pulse lengths of about 10\,ns have been measured
for a large range of amplitudes, which is needed for a good quality of
the single-pixel-spectra. In order to minimize the number of
calibration runs in test beam operations, it is important to have a
reasonably uniform light output for a large number of channels. First
tests are ongoing to investigate, if this can be achieved by closing
solder jumpers to add the bias capacitors C$_2=22$\,pF and
C$_3=82$\,pF to the default capacitor C$_1$. More details of the
driver circuit can be found in~\cite{EPT}. An alternative concept is
also studied that is based on few strong LEDs on special interface
boards, where the light is distributed via notched fibers~\cite{Jiri}.

As shown in Fig.~\ref{fig:slab}, each slab is connected to a data
acquisition system (DAQ). The current setup comprises a Central
Interface Board that hosts the Detector Interface (left), the steering
board for the calibration system (middle) and the power module
(right), that distributes all voltages needed in the slab. More
details about the concept of the DAQ are reported in~\cite{EPT} and
references therein.

\section{Measurements and results} \label{sec:results}

The main goal of the current tests of the prototype modules is the
commissioning of the design concept for a multi-channel prototype. On
one side, this requires tests of the functionality and performance of
all subcomponents in the laboratory as well as in a test beam
environment. This has been done extensively in the past
(see~\cite{EPT} and references therein) and is still ongoing. Results
of recent tests are reported in the following. On the other side,
system aspects have to be considered and the performance of larger
setups as shown in Fig.~\ref{fig:slab} has to be tested. This will be
possible with the next, final stage of the prototype operation, when a
complete layer with up to 18 HBUs (72 ASICs) can be operated and read
out.

\subsection{Time measurement} \label{sec:TDC}

\begin{figure}
\centering
\subfigure[] {\includegraphics[width=2.4in, height=1.5in]{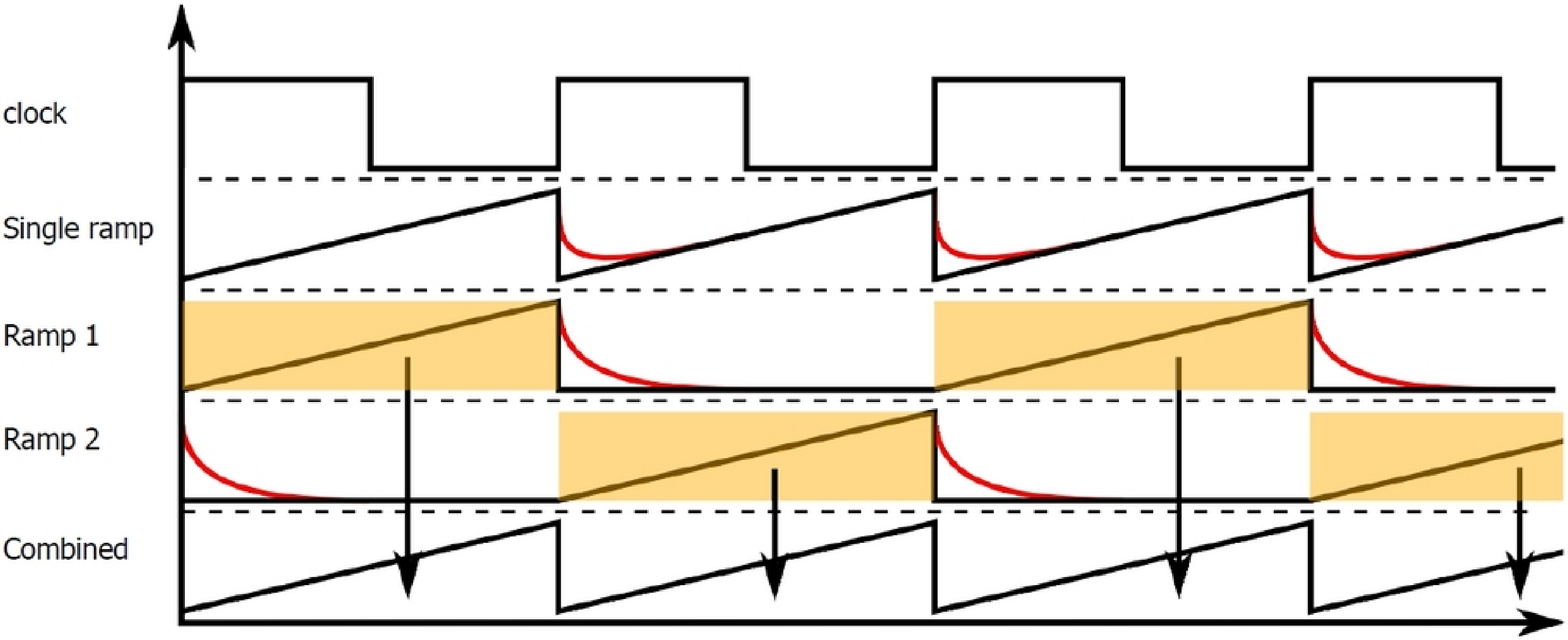}\label{fig:TDC}}
\hspace{1.5cm}
\subfigure[] {\includegraphics[width=2.4in]{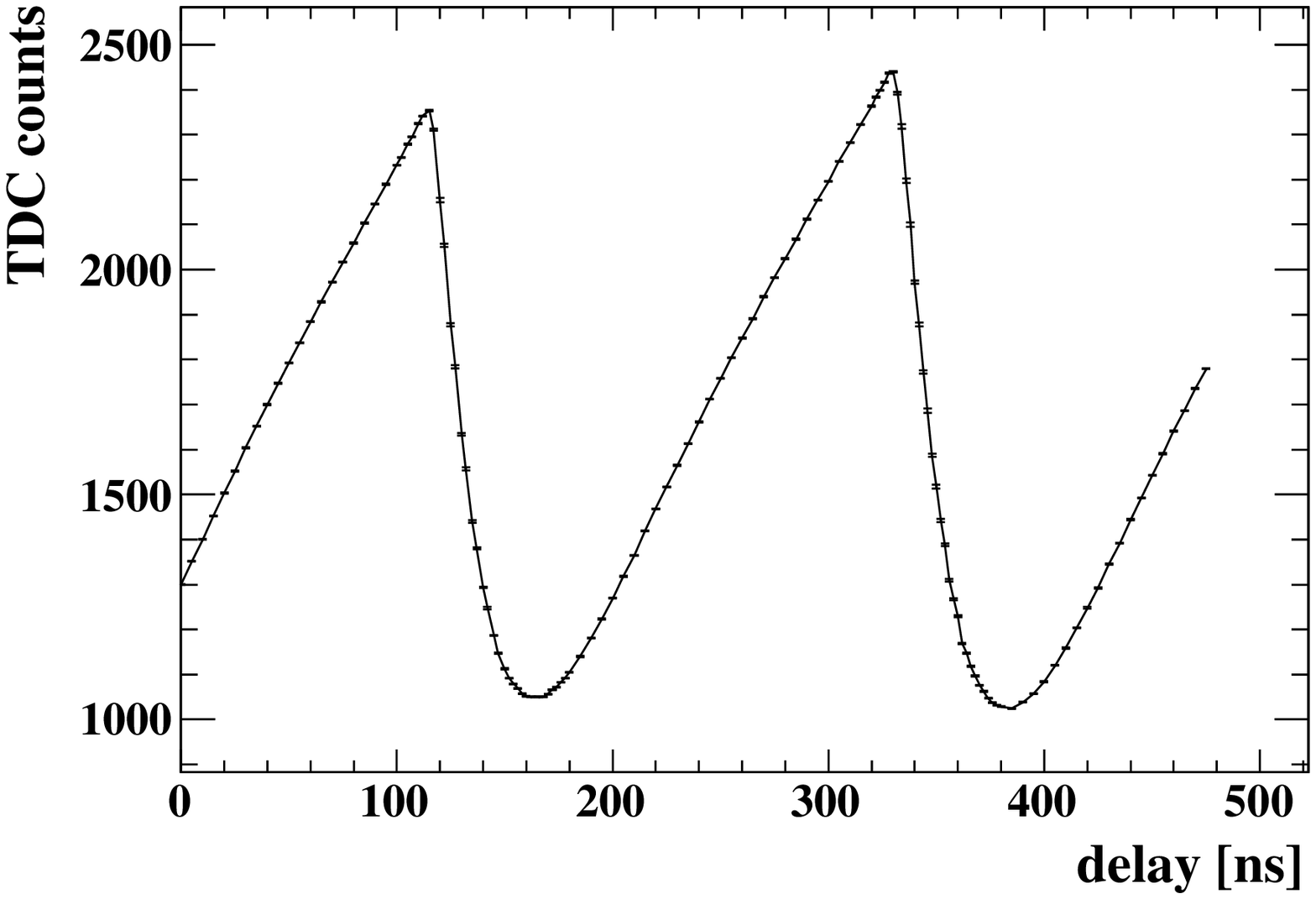}\label{fig:LCramp}}
\caption{(a) Schematic working principle of the dual slope TDC. Reset
  effects are shown in red. (b) TDC ramp in LC mode.}
\end{figure}

In addition to the energy measurement of the shower, it is useful to
measure also the arrival time of a signal in each cell. With this
information one can distinguish between prompt and delayed shower
components, which can be used e.g. for the identification of late
neutrons. This discrimination improves the performance of the particle
flow algorithm. The SPIROC2b ASIC has the capability of measuring the
arrival time of a signal relative to the bunch clock (5\,MHz at a LC)
with a 12-bit Wilkinson TDC. The voltage ramp is started with the
rising edge of the clock and is reset with the next rising clock edge.
Since the reset of the ramp introduces dead time, two ramps have been
implemented. The first ramp is only active on even clock cycles, while
the second ramp is only active on odd clock cycles. A multiplexer
switches between these two ramps (which again may introduce some dead
time). The working principle of the time measurement is depicted in
Fig.~\ref{fig:TDC}.

There are two possible operation modes foreseen in the current HBU
design. In the LC mode the ramp has a length of 200\,ns (to match the
bunch structure of the machine), while in test beam mode the ramp has
a length of about 5\,$\mu$s (in order to reduce dead time due to
multiplexing). These values can be further adapted to the test beam
needs in the DAQ firmware. To exploit the full dynamic range of the
TDC in case of a change of the ramp length, the slope of the ramp can
also be changed by SPIROC2b bias points. The physics goal of the
performance of the time measurement is to achieve a resolution of
about 1\,ns - 3\,ns in test beam mode to be able to distinguish
between prompt and late shower components. In order to test the
performance of the TDC, a signal was injected into one input channel
and the ramp was measured with a stepwise delay of the injected signal
with respect to the SPIROC2b clock. Figure~\ref{fig:LCramp} shows a
typical TDC ramp measured in LC mode. The two ramps have slightly
different peak values, while the dead time between the maximum of one
ramp and the start of the other ramp is caused by the multiplexer.
Both issues will be improved in the next generation of the ASIC. The
resolution is determined to be $\sim$300\,ps. This can be improved in
the future by optimizing the ramp slope and exploiting the full
dynamic range of the TDC. The measurement also shows that the use of
the LC mode in a test beam environment is not ideal, since about 50\%
of a clock cycle would be dead time. Therefore, the clock period is
tuned to about 5\,$\mu$s. This compromises the resolution, since the
slope of the TDC ramp has to be chosen less steep. The measured ramp
in test beam mode provides a resolution of about 3\,ns. This can also
be improved by tuning the TDC ramp as well as optimizing the clock
period, such that a final resolution of 1\,ns will be achievable in
test beam mode.

\subsection{Light amplitude equalization of LED system and SiPM saturation}

\begin{figure}
\centering
\subfigure[] {\includegraphics[width=2.3in]{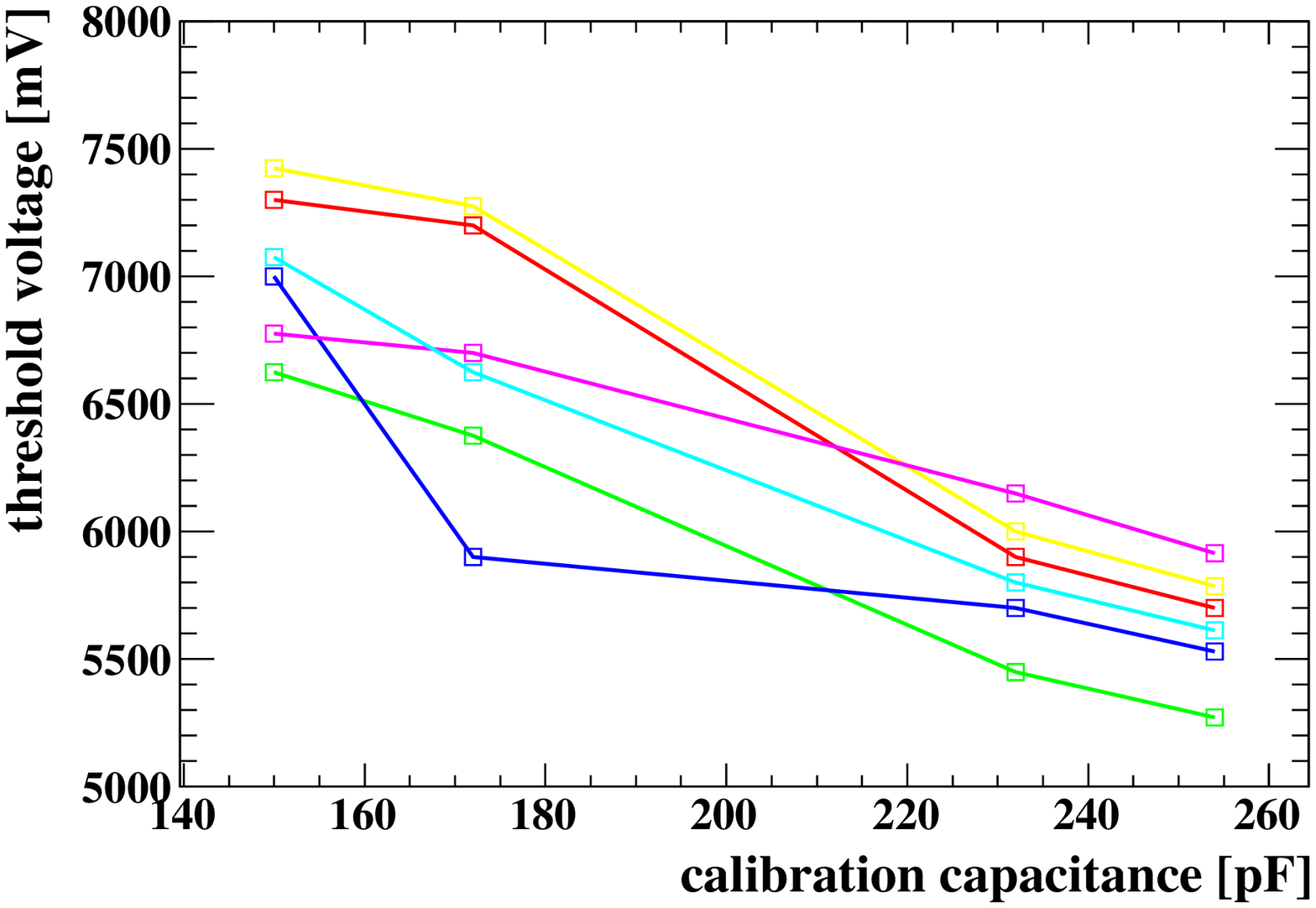}\label{fig:LED_threshold}}
\hspace{1.5cm}
\subfigure[] {\includegraphics[width=2.3in]{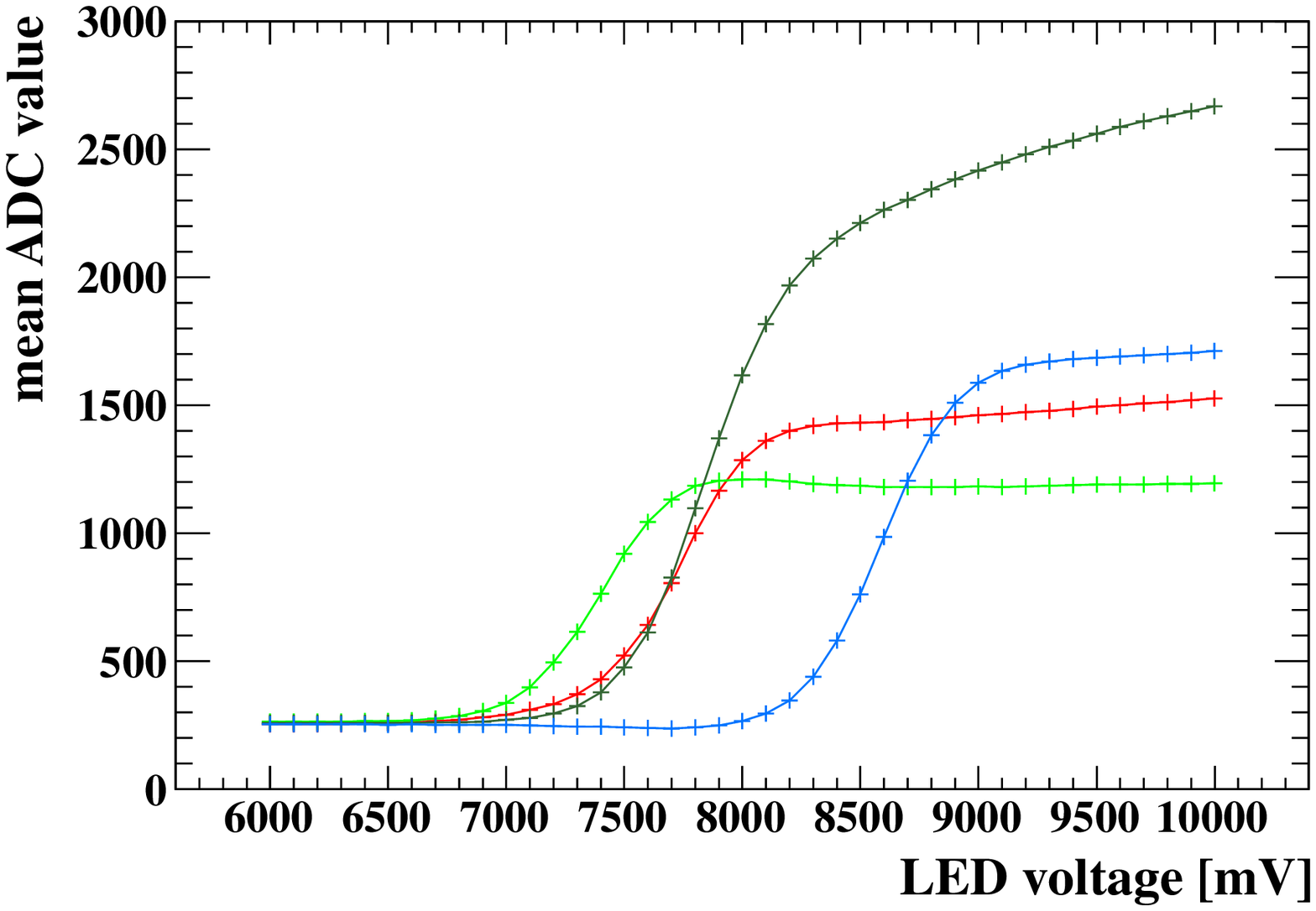}\label{fig:SiPM_saturation}}
\caption{(a) Dependence of the LED bias voltage where the LED starts
  to emit light from the bias capacitance. (b) SiPM saturation
  measured with the integrated LED calibration system.}
\end{figure}

The main goal of the LED calibration system is to measure the gain of
each SiPM by pulsing LED light of low intensity into the tiles. Since
this has to be achieved in as few calibration runs (or with as few
different LED bias voltages) as possible, the light output of the LEDs
have to be equalized, as discussed in Sec.~\ref{sec:status}. The goal
is to find a suitable combination and optimal values of the capacitors
C$_1$, C$_2$ and C$_3$ for each tile, such that the range of the LED
bias voltage that is needed for the LED to produce photons is as small
as possible. Each possible combination of capacitors has been used to
measure the LED light output as a function of the bias voltage for a
small set of tiles. Figure~\ref{fig:LED_threshold} shows the bias
voltage, where the measured LED light amplitude exceeds a small
threshold (starts flashing), as a function of the total capacitance
(sum of the nominal values of the capacitors). As expected, the
threshold decreases as the capacitance increases. The range of the
bias voltage can already been decreased with this method.
Nevertheless, further optimization of the capacity values is needed
and further studies will be done in order to optimize the calibration
procedure.

The second task of the LED system is to measure the saturation of the
SiPMs. Figure~\ref{fig:SiPM_saturation} shows some typical saturation
curves for different channels measured with LED light in the low gain
mode of the ASIC. Note that the visible saturation is not due to ADC
saturation, since the dynamic range of 12-bit is not fully exploited.

\subsection{MIP and light yield test beam measurement}

\begin{wrapfigure}{r}{0.4\columnwidth}
\centerline{\includegraphics[width=0.35\columnwidth]{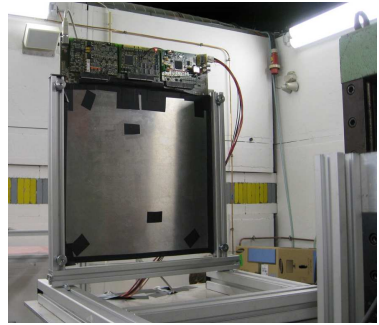}}
\caption{Setup of the light-tight HBU cassette as it is mounted on a
  movable stage at the DESY test beam facility. On top of the cassette
  the detector interface modules are visible.}
\label{fig:TB}
\end{wrapfigure}

One HBU that is equipped with scintillator tiles is currently under
test at the DESY test beam facility. Electrons with an energy of
2\,GeV are used to investigate the response of the system to MIPs. For
this purpose the HBU is enclosed into a light-tight aluminum cassette
and mounted on a movable stage in order to scan all channels (see
Fig.~\ref{fig:TB}). The MIP signals are measured in auto-trigger mode,
where the threshold is optimized for each channel to measure a full
MIP spectrum, while suppressing most of the SiPM noise. Since in
SPIROC2b the triggers of all channels are connected with a logical OR,
the preamplifiers of all channels that are not used are switched off.
The measurements have been done in high gain mode with a preamplifier
feedback capacitance of 100\,fF and a shaping time of 50\,ns.

\begin{figure}
\centering
\subfigure[] {\includegraphics[width=2.3in]{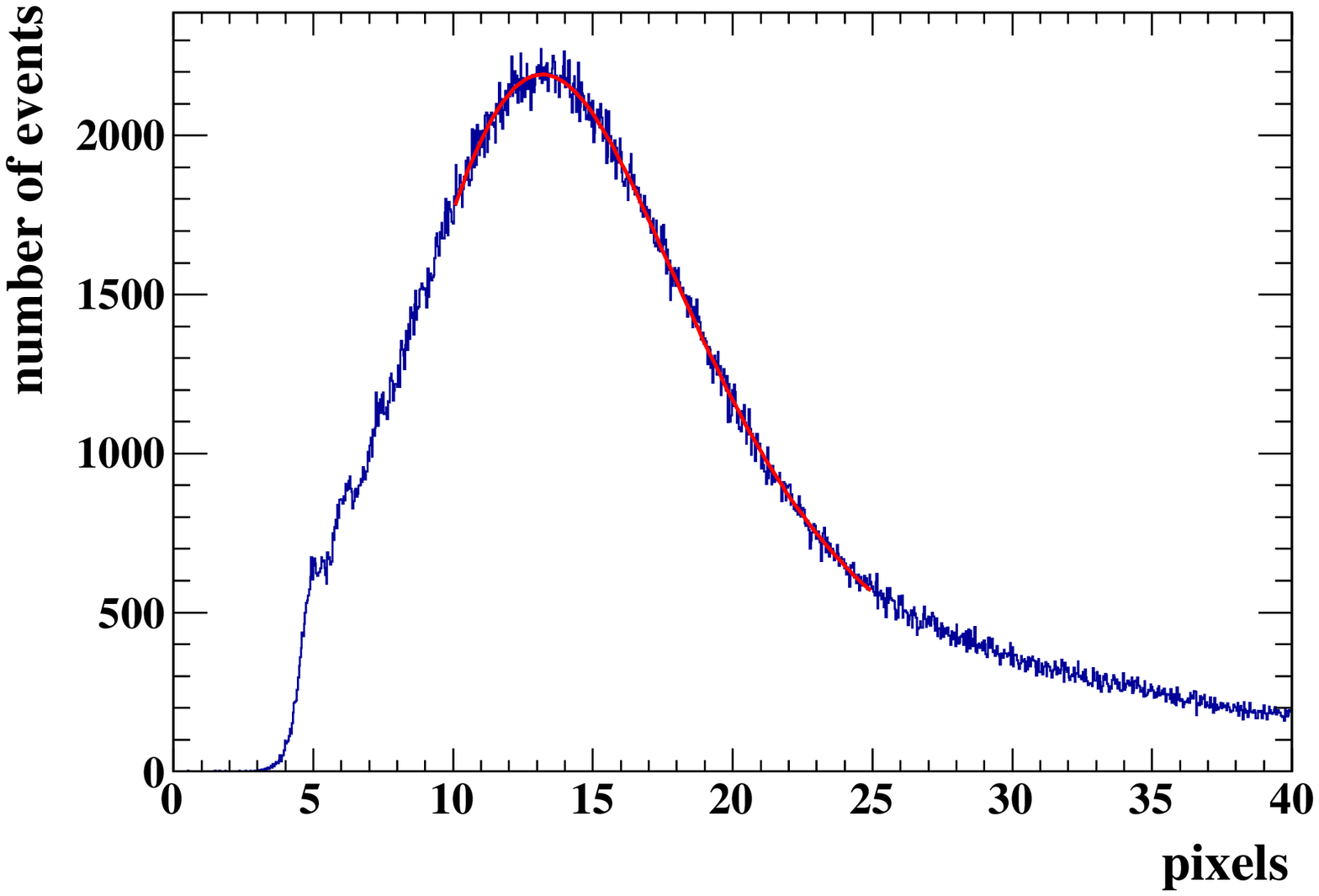}\label{fig:MIP}}
\hspace{1.5cm}
\subfigure[] {\includegraphics[width=2.3in]{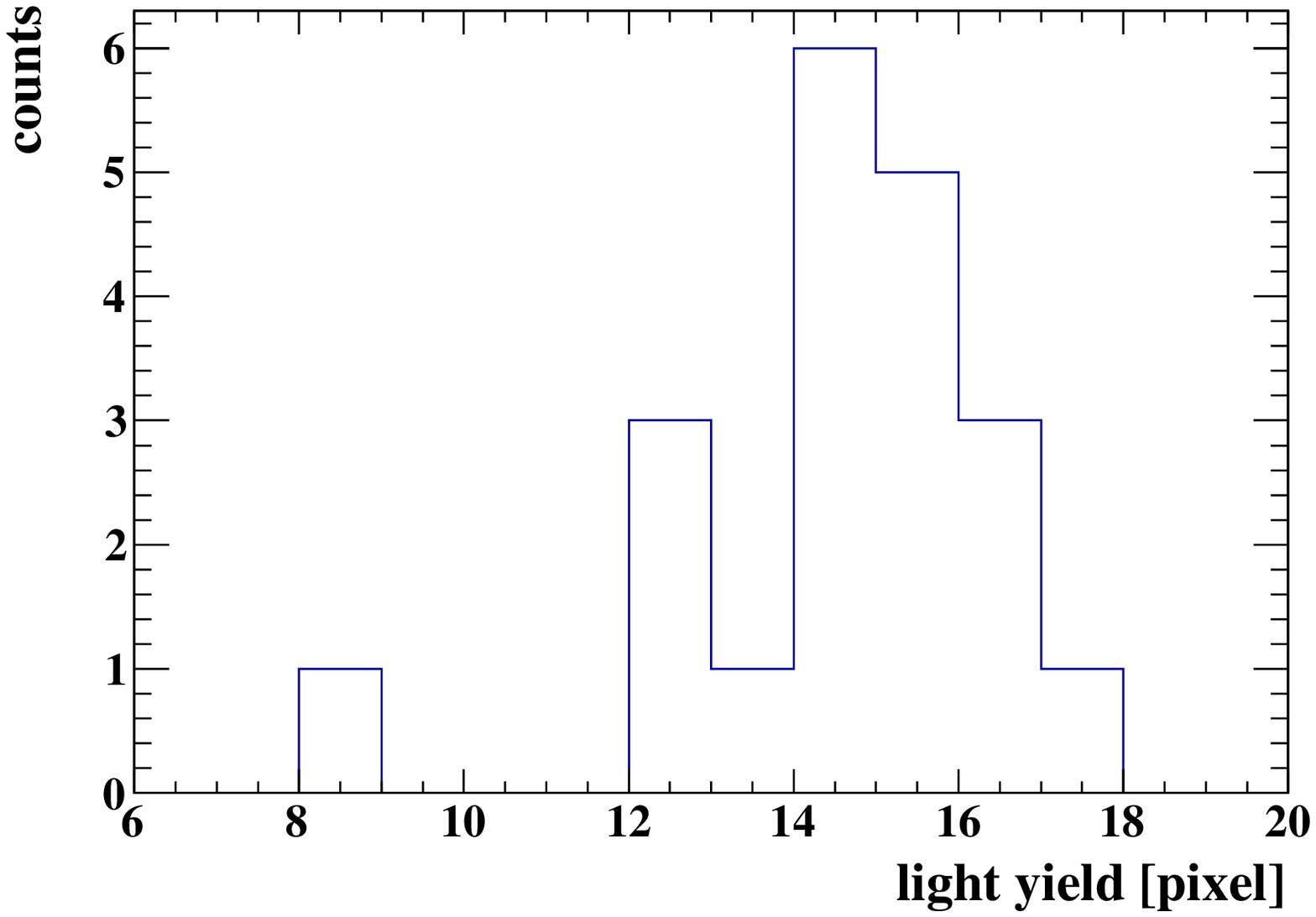}\label{fig:LY}}
\caption{(a) Typical MIP spectrum obtained from a 2\,GeV electron beam
  and (b) light yield distribution measured with the latest HBU
  version at the DESY test beam.}
\end{figure}

Figure~\ref{fig:MIP} shows a typical MIP spectrum. The measured ADC
value from the front-end electronics is converted into a pixel number
by using the measured gain from single-pixel-spectra taken with LED
light. The most probable value is around 15 pixels. Also the trigger
efficiency curve can be observed at the beginning of the spectrum. The
spectrum is fitted with a Landau function convolved with a Gaussian
function to determine the exact position of the most probable value
(light yield). The distribution of the light yield of all investigated
tiles is shown in Fig.~\ref{fig:LY}. The mean value is at 15 pixels,
which is the design value for the used tiles.

\section{Summary and outlook}

A new engineering prototype for an analog hadron calorimeter is
currently being developed by the CALICE collaboration. The goal is to
show that a realistic LC detector with fully integrated front-end
electronics can be built. The main challenge for the near future is to
construct a full LC detector layer to test the signal integrity and
the concept of power pulsing. Furthermore, a layer for operation in a
hadron test beam environment will be built and used for measuring the
time evolution of hadronic showers.

In order to achieve these goals, the calorimeter base units are tested
in the laboratory as well as in the DESY test beam to characterize
their features and to test the functionality and performance of all
subcomponents as well as overall system aspects. Some of the recent
measurement results are shown in this report, including the time
measurement performance of the ASIC, some aspects of the LED
calibration system and latest test beam measurements of the light
yield of the new scintillator tiles.

The next important steps are the construction of multi-HBU setups to
further investigate the integrated read-out electronics and the
development of a scalable data acquisition system.

\section*{Acknowledgments}

The authors gratefully thank Karsten Gadow, Erika Garutti, Peter
G\"ottlicher, Benjamin Hermberg, Mathias Reinecke, Julian Sauer, Felix
Sefkow and Sebastian Weber for very useful discussions and valuable
contributions to the results presented here.


\begin{footnotesize}


\end{footnotesize}


\end{document}